\documentclass[preprint]{elsarticle}
\usepackage{amsmath}
\usepackage{amssymb}
\usepackage{epsfig}
\usepackage{epstopdf}
\usepackage[latin5]{inputenc}
\usepackage{subfigure}
\usepackage{lineno,hyperref}

\modulolinenumbers[5]

\usepackage{color}

\numberwithin{equation}{section}

\begin{document}

\title{On the time shift phenomena in epidemic  models }

\author[a]{Ayse Peker Dobie}
\ead{pdobie@itu.edu.tr}

\author[a]{Ali Demirci\corref{cor1}}
\ead{demircial@itu.edu.tr}

\author[b]{Ayse Humeyra Bilge }
\ead{ayse.bilge@khas.edu.tr}

\author[a]{Semra Ahmetolan}
\ead{ahmetola@itu.edu.tr}

\cortext[cor1]{Corresponding author}
\address[a]{Department of Mathematics, Faculty of Science and Letters, Istanbul Technical University, Istanbul, Turkey}
\address[b]{Department of Industrial Engineering, Faculty of Engineering and Natural Sciences,  Kadir Has  University, Istanbul, Turkey}

\begin{abstract}
In the  standard Susceptible-Infected-Removed (SIR) and  Susceptible-Exposed-Infected-Removed (SEIR) models, the peak of infected individuals coincides with the inflection point of removed individuals.  Nevertheless, a  survey based on the data of the 2009 H1N1 epidemic in Istanbul, Turkey \cite{bib19}  displayed  an unexpected time shift  between the hospital referrals and fatalities. With the motivation of investigating the underlying reason, we use multistage $SIR$ and $SEIR$ models to provide an explanation for this  time shift.  Numerical solutions of these  models present strong evidences that the delay is approximately half of the infection period of the epidemic disease. In addition,  graphs of the classical $SIR$  and the multistage $SIR$  models; and the classical $SEIR$  and the multistage $SEIR$  models are compared for various epidemic parameters. Depending on the number of stages, it is observed that the delay varies for relatively small stage numbers whereas it does not change for large numbers in multistage systems. One important result that follows immediately from this observation is that this fixed delay for large numbers explains the time shift.   Additionally, depending on the stage number and the duration of the epidemic disease, the distance between the points where each infectious stage reaches its maximum is  found approximately both graphically and qualitatively for both  systems. Variations of the time shift, the maximum point of the sum of all infectious stages, and the inflection point of the removed stage are observed subject to the stage number $N$ and it is shown that these variations stay unchanged for greater values of $N$.

\end{abstract}


\date{\today}

\maketitle

\section{Introduction}

From the early attempts~\cite{bib1}~\cite{bib2}~\cite{bib3}~\cite{bib4}  to recent studies, epidemic modeling which is applicable in a wide range of fields from informatics~\cite{bib5}~\cite{bib6} to chemistry~\cite{bib7}~\cite{bib8}~\cite{bib9} has drawn the attention of researchers in various disciplines. Since the basic compartmental  model Susceptible-Infected-Removed  ($SIR$) which  is commonly used to model diseases for which the infection confers permanent immunity was introduced by Kermack and McKendrick in 1927~\cite{bib4}, other compartmental models ~\cite{bib10}~\cite{bib11}~\cite{bib12} have been developed to model diseases with different structures and dynamics. Especially in recent years,  major outbreaks such as avian flu in 2005, swine influenza in 2006 and H1N1 influenza in 2009 have highlighted the need for more effective and reliable models to control the spread of disease and to provide a better knowledge for the prediction of future threats and for the development of stronger containment strategies.  In the works   \cite{bib13}-\cite{bib18}, some results on the modelling of different types of epidemic diseases, the solution form of these models, the observation of global stability and the determination of the final size of the epidemic are obtained. In\cite{bib13}, the $SEIR$ model which differs from the classical $SIR$ model by  the introduction  of an exposed class $E$ is investigated and a criterion on the existence of the global stability  with respect to the basic reproduction number $R_0$ is developed. In \cite{bib15}, some similar criteria on global stability are derived for the   $SEIS$ model which can be regarded as a model with no immunity, and  different  $SIR$ models are examined in \cite {bib16},\cite{bib17}. Moreover, the works \cite{bib17},\cite{bib18} provide some useful results on the final size of the epidemic for   $SIR$ models. With the same motivation of these works but rather  a different contribution to literature, we  use a multistage   model in this article to explain the time shift observed in several surveys.

In the literature, available observed data that is used to construct the ordinary differential equation system representing the classical $SIR$  epidemic model is based on the curve of removed individuals. Usually,  this curve is obtained by taking into consideration only the  fatality data of the epidemic disease whereas  in some researches, not only the fatality data  but also the hospitalization data for the epidemic are  taken into account in the construction of the system. In some of these researches, it is  shown that there exists a delay between the peak of  the hospitalization (infectious) curve and the inflection point of the fatality (removed) curve based on the  data collected. The original contribution of this article to literature is that we  explain this time shift by the multidimensional form of  $SIR$ and $SEIR$ models and also show that the expected delay is approximately half of the infectious period of the epidemic disease for both of the multistage systems.

In the second section of this work, we give a brief summary of the classical epidemic $SIR$ and $SEIR$  models  and define the multistage form of these models that will be used in further analysis. The graphs obtained by the numerical evaluations of the classical $SIR$  and $SEIR$ models and their multistage forms are also given. Analysis of these graphs confirms that  the multistage $SIR$ and $SEIR$ models explain  the time shift observed  in several surveys. In the third section,  the distance between the points where successive stages and hence any two stages assume their maximum is  found approximately.  In the fourth section, the evaluation of delay for different epidemic parameters is presented by using the numerical evaluations of these multistage  models.  The last section includes a summary of the  results obtained in the previous sections  as well as information for future analysis.

\section{Standard epidemic models and epidemic models with multiple infectious stages}
\label{Standard epidemic models and epidemic models with multiple infectious stages}

The SIR model also known as the Kermack-McKendrick model is commonly used to model diseases for which the removed individuals are assumed to be immune to reinfection. In addition, the total population with constant size is divided into three
distinct compartments the size of which change with time $t$. These compartments are called the susceptible class $S$, the
infective class $I$ and the removed class $R$. Individuals who are able to become infected are members of class $S$ until they are infected with a pathogen and become capable of transmitting the disease to others. Then, they move from the class $S$ into the class $I$ once they are
infected and then from $I$ to $R$ once they  recover or die. Childhood illnesses like measles or rubella  are good examples for $SIR$ model.

The  SIR  epidemic model without vital dynamics; that is, the recruitment of new susceptible through birth or immigration as well as the loss through mortality or emigration  are ignored, is defined by the following system of nonlinear  ordinary differential equations
\begin{eqnarray}
\label{eq:schemeP}
\begin{aligned}
SIR:~~~~S'&=-\beta SI\\
I'&=\beta SI-\gamma I\\
R'&=\gamma I
\end{aligned}
\end{eqnarray}
where the coefficient $\beta$ refers to the disease transmission rate and ${1}/{\gamma}$ refers to the duration of infection period. Note that since $S'+I'+R'=0$,  we may assume  $S+I+R=1$ by the use of appropriate normalization.

The standard $SIR$ model ignores a latent phase which is the delay between the
time of the acquisition of infection and the onset of infectiousness. In order to define this latent phase, the introduction to the $SIR$ model of an exposed class $E$ whose members are individuals who have been infected with a pathogen but are not yet infectious due to the incubation period of pathogen
 yields the $SEIR$ model.  Chicken pox is suitable for the $SEIR$ model which is defined by the following system of  ordinary differential equations
\begin{eqnarray}
\label{eq:schemeP}
\begin{aligned}
 SEIR:~~~~
S'&=-\beta SI\\
E'&=\beta SI-\epsilon E\\
I'&=\epsilon E-\gamma I\\
R'&=\gamma I
\end{aligned}
\end{eqnarray}
where ${1}/{\epsilon}$ is the mean exposed period. Note that  we may  assume  $S+E+I+R=1$ by the use of appropriate normalization.

\begin{figure}[ht]
\includegraphics[scale=0.42]{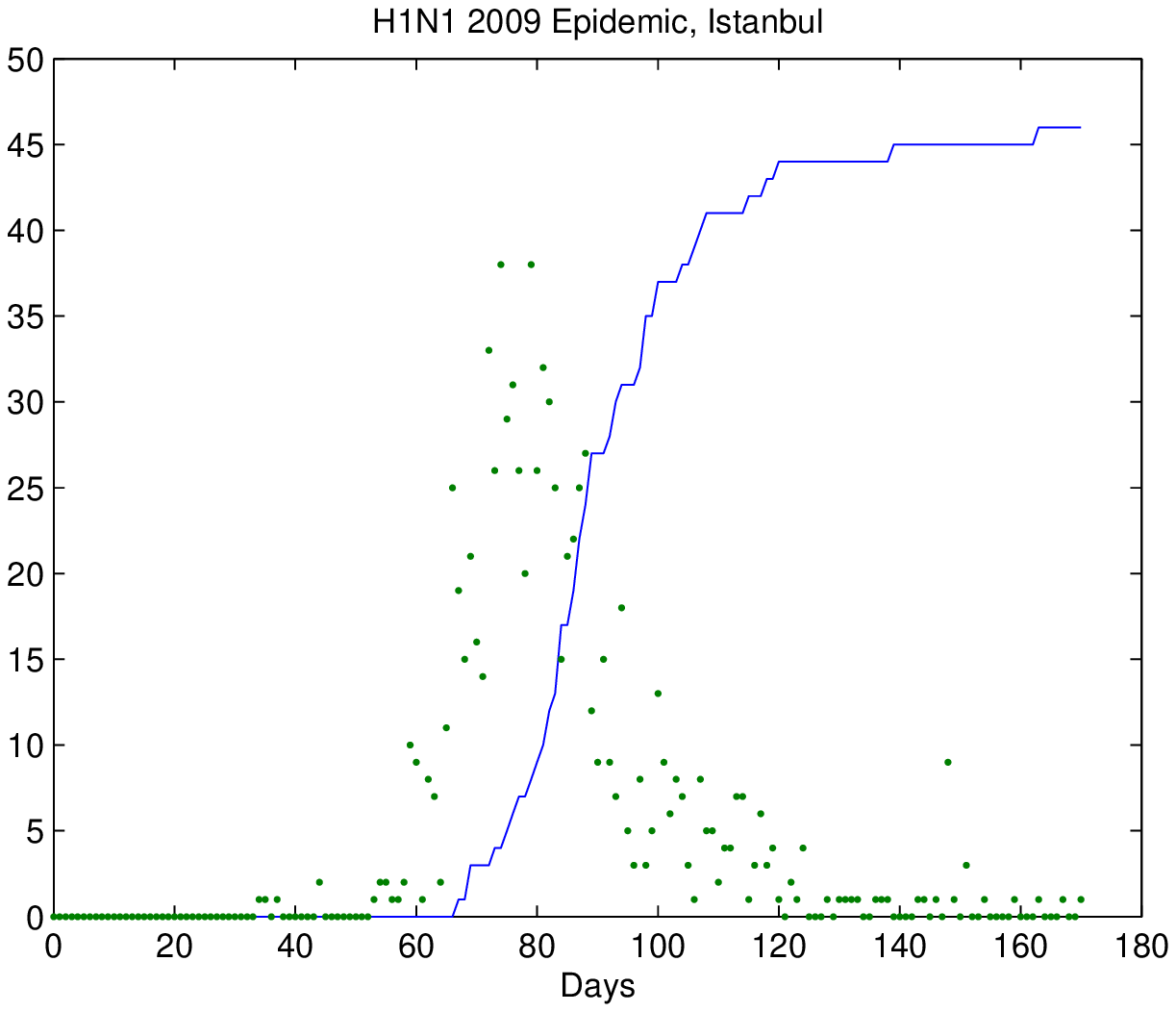}
\includegraphics[scale=0.42]{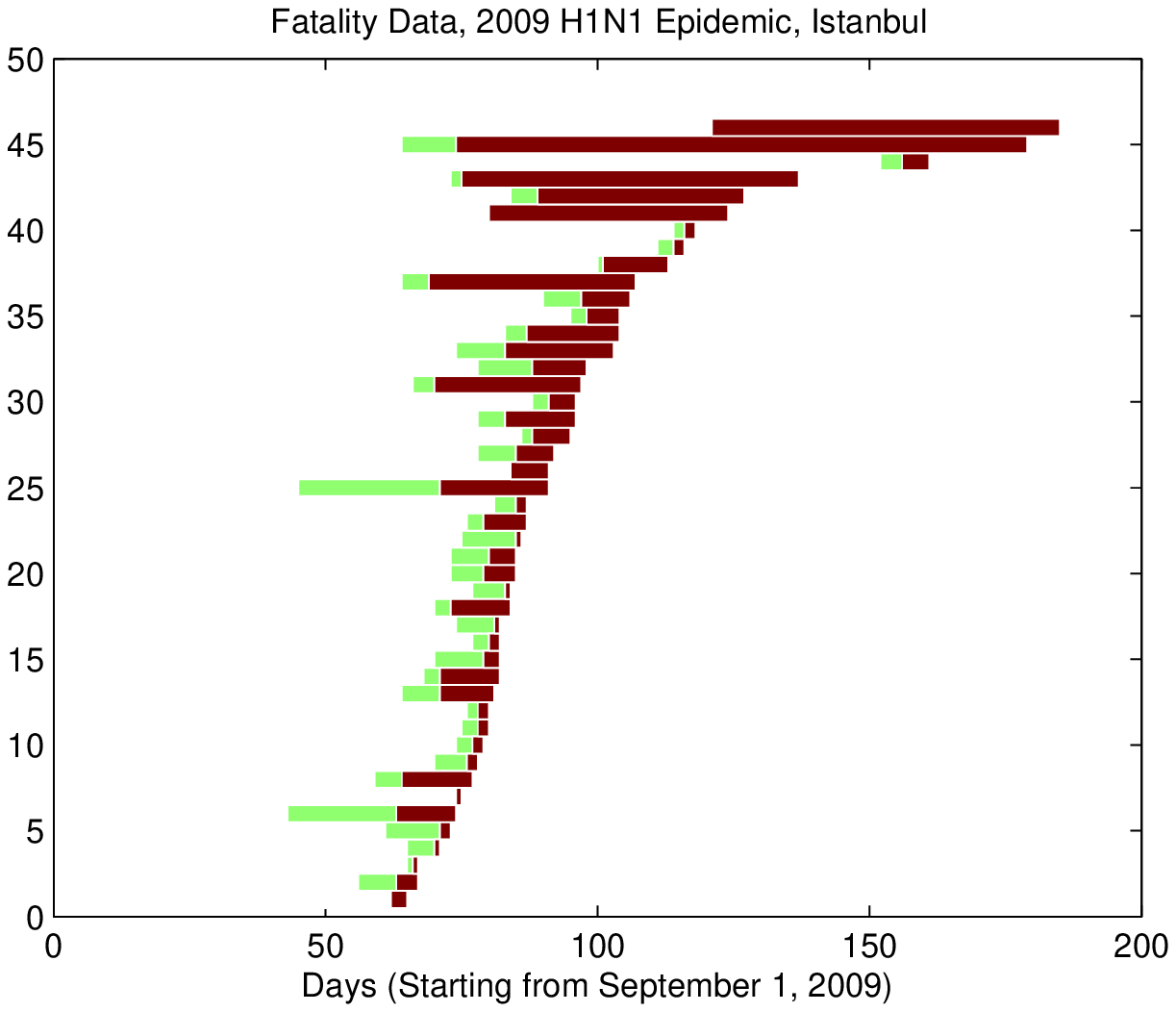}
\caption{The daily number of referrals to hospitals and cumulative number of fatalities for the 2009 H1N1
epidemic in Istanbul, Turkey. The scatter in the number of daily referral to hospitals indicates that the
hospitalization rate varies during the epidemic. The asymmetry of the incidence curve is still observable.
The fatalities shown here are adjusted to at most 15 days after referral to the hospital (left panel). The duration
of symptoms prior to hospitalization (light color) and the duration of hospitalization prior to death (dark
color) for the fatalities in Istanbul, Turkey during due to 2009 H1N1 epidemic (right panel)}
\end{figure}

Although these two models are suitable for mathematical modelling of seasonal diseases, an unexpected time shift which is shown in Fig 1a  was observed in the modelling of the 2009 H1N1 epidemic in Istanbul, Turkey \cite{bib19}. As it can be seen in Fig 1b, the fatality data as well as the hospitalization data were collected in the construction of the system.
In this study,  we also use multiple infectivity periods similar to \cite{bib13},\cite{bib14} to explain this unforeseen delay. The approach of multiple infectious stages consists of replacing the single infectious stage $I$ with $N+1$ substages denoted by $I_i$ which is the density of individuals in the $i$th infectious stage. Each of these stages may have   different infectivity $\beta_i$ and may have a variable infectious period  $1/\gamma_i$
satisfying the condition for the $SIR$ model
$$
\frac{1}{\gamma_0}+\frac{1}{\gamma_1}+\cdots+\frac{1}{\gamma_N}=\frac{1}{\gamma}
$$
and for the $SEIR$ model
$$
\frac{1}{\gamma_1}+\frac{1}{\gamma_2}+\cdots+\frac{1}{\gamma_N}=\frac{1}{\gamma}.
$$

Then the multistage $SIR$ and $SEIR$ epidemic models are defined by the following systems
\begin{eqnarray}
\label{eq:schemeP}
\begin{aligned}
Multistage~~ SIR:~~~~S'&=-S (\beta_0 I_0+\beta_1 I_1+\cdots +\beta_N I_n)~~~~~~~~~~~~~~~~\\
I'_0&=S (\beta_0 I_0+\beta_1 I_1+\cdots +\beta_n I_N)-\gamma_0 I_0\\
I'_1&=\gamma_0 I_0-\gamma_1 I_1\\
\cdots&\cdots\cdots\\
\cdots&\cdots\cdots\\
I'_N&=\gamma_{N-1} I_{N-1}-\gamma_N I_N\\
R'&=\gamma_N I_N
\end{aligned}
\end{eqnarray}

\begin{eqnarray}
\label{eq:schemeP}
\begin{aligned}
 Multistage~~SEIR:~~~~
S'&=-S (\beta_1 I_1+\cdots +\beta_ NI_N)~~~~~~~~~~~~~~~~~~~~~~~~~~~~\\
E'&=S(\beta_1 I_1+\cdots +\beta_N I_N)-\epsilon E~~~~\\
I'_1&=\epsilon E-\gamma_1 I_1\\
I'_2&=\gamma_1 I_1-\gamma_2 I_2\\
\cdots&\cdots\cdots\\
\cdots&\cdots\cdots\\
I'_N&=\gamma_{N-1} I_{N-1}-\gamma_N I_N\\
R'&=\gamma_N I_N.
\end{aligned}
\end{eqnarray}

The linear parts of apparently different infectious stages for $i\geq 1$ in the multistage $SIR$ model and $i\geq 2$ in the multistage $SEIR$ model have a similar structure. To facilitate this,
we write the linear parts of each equation above as  a system and then rearrange and rename as follows to keep the models as  clear and simple  as possible
\begin{eqnarray}
\label{eq:schemeP}
\begin{aligned}
SJR:~~~~J&= I_0+\frac{\beta_1}{\beta_0} I_1+\cdots +\frac{\beta_N}{\beta_0} I_N\\
S'&=-\beta_0 SJ\\
I'_0&=\beta_0 SJ-\gamma_0 I_0\\
I'_i&=\gamma_{i-1} I_{i-1}-\gamma_{i} I_{i},\,\,\,for\,\,\,i=1,....N\\
R'&=\gamma_{N}I_N
 \end{aligned}
\end{eqnarray}

\begin{eqnarray}
\label{eq:schemeP}
\begin{aligned}
SEJR:~~~~J&=I_1+\frac{\beta_2}{\beta_1} I_2\cdots +\frac{\beta_N}{\beta_1} I_N\\
S'&=-\beta_1 SJ\\
E'&=\beta_1 SJ-\epsilon E\\
I'_1&=\epsilon E-\gamma_1 I_1\\
I'_i&=\gamma_{i-1} I_{i-1}-\gamma_{i} I_{i},\,\,\,for\,\,\,i=2,....N\\
R'&=\gamma_{N}I_N.
\end{aligned}
\end{eqnarray}

Subsequently, the numerical evaluations of these two systems, $SJR$ and $SEJR$ defined above will be used for some of the structural comparisons of the classical models and multistage models. Furthermore, Matlab\,\textsuperscript{\tiny\textregistered} ODE45 solver is used for all the numerical evaluations for which the following initial conditions are valid for the $SJR$ model
\begin{equation}
S(0)=1-10^{-4},\,\,I_{0}(0)=10^{-4},\,\,\,I_{i}(0)=0,\,\,\,i=1,....,N,\,\,\,\,R(0)=0
\end{equation}
and for the $SEJR$ model
\begin{equation}
S(0)=1-10^{-4},\,\,E(0)=5*10^{-5},\,\,I_{1}(0)=5*10^{-5},\,\,I_{i}(0)=0,\,\,i=2,....,N,\,\,R(0)=0.
\end{equation}

Initially, numerical evaluations of the classical $SIR$ and $SEIR$ models are made for specifically chosen epidemic parameters and the results are given in Fig2. For the evaluations of the classical models, $\gamma$ and $\epsilon$ are fixed as $1/5$ and $1/3$, respectively  whereas the value of $R_0$ is chosen to be  $2.5$ and $10$. For all of these specific values, the maximum of the infectious stage and the inflection point of the removed stage occur at the same point in time. Therefore, the time shift does not conform to classical models.
\begin{figure}[ht]
\includegraphics[scale=0.57]{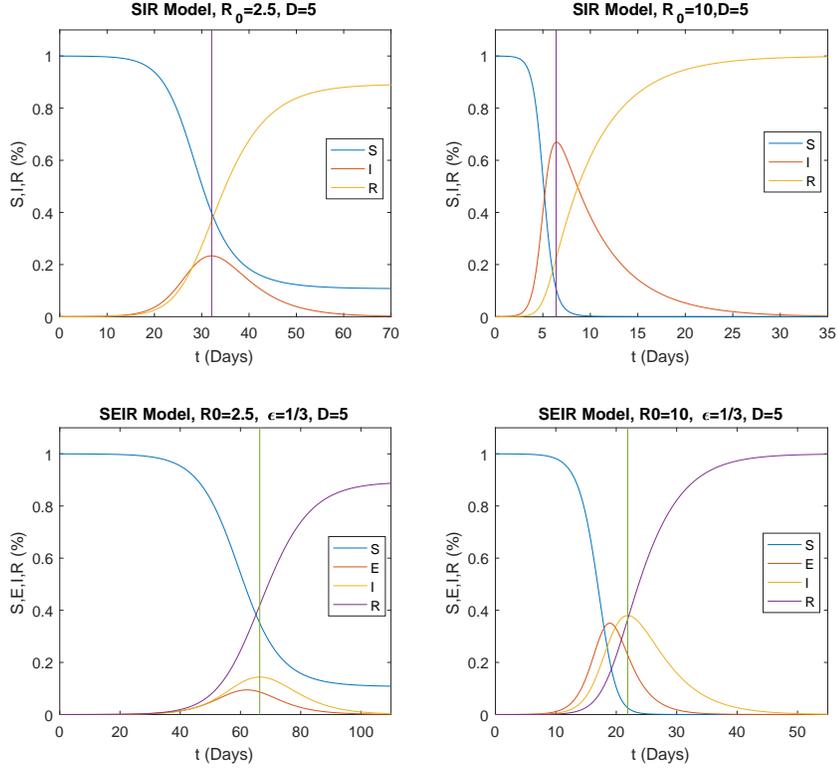}
\caption{Numerical solutions of classical SIR (1) and SEIR (2) models with $R_0=2.5$ and $10$. Here $\epsilon$ and $\gamma$ are chosen as $1/3$ and $1/5$ ($D=5$), respectively. Time-shift does not occur in classical models. }
\end{figure}

Later, numerical evaluations of the  multistage $SIR$ model are obtained for various stage numbers. Firstly, $R_0$ is set as $2.5$ while the stage number $N$ is given the values $5, 10, 30$ and $60$, and the corresponding graphs of the solutions are shown in Fig3. In these evaluations, $\gamma_i$ is chosen to be $N*0.2$ for  $i=0,1,...,N$ where $N$ is the infectious stage number.  In Fig3, the difference between the maximum point of the curve representing the sum of the infectious stages and the inflection point of the  removed curve can clearly be seen. As predicted, the system given by  (5) confirms the delay and therefore seems adequate to explain the unexpected  time-shift between infectious and removed stages observed in Istanbul data \cite{bib19}. However, another analysis of Fig3 shows that the curves of $SJR$ model differ from the curves of classical $SIR$ model. The underlying reason for this difference is that the classical $SIR$ system does not depend on the stage number and hence the value of $\gamma$ stays fixed, whereas for the multistage $SIR$ model, the values of $\gamma_i$ change according to the stage number $N$.

\begin{figure}[ht]
	\includegraphics[scale=0.57]{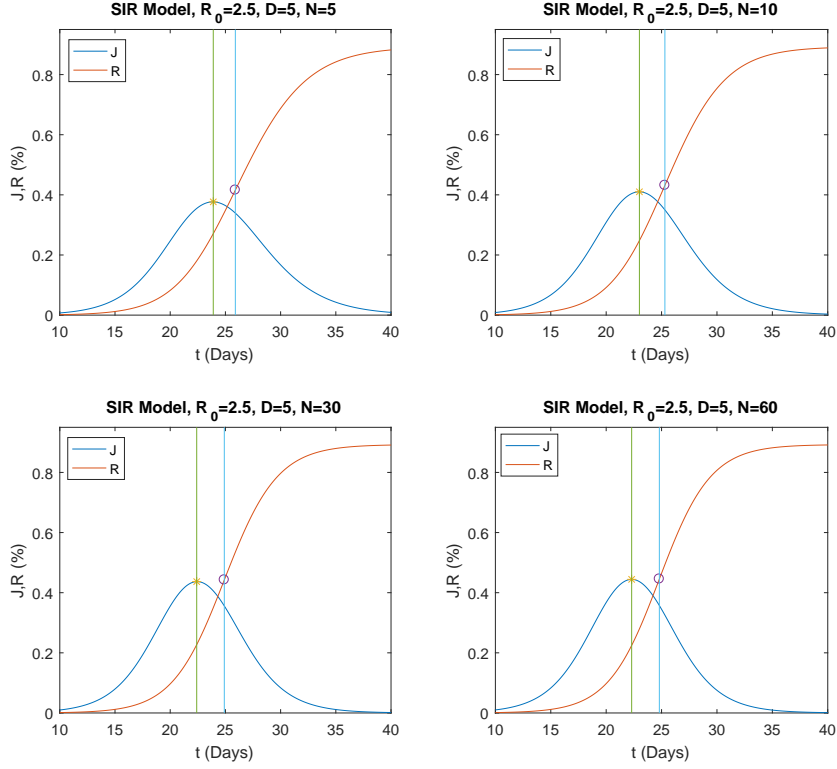}
	\caption{Numerical solutions of multistage SIR model for different stage numbers $N=5,10, 30$ and $60$. Here $R_0=2.5$ and $\gamma=N$x$0.2$. In the graphs, $*$ represents the location of maximum value of $J$ and $o$ represents the location of the inflection point of $R$. Graphs show that there is a time-shift between these points.  }
\end{figure}

Then, numerical evaluations of the  multistage $SEIR$ model are obtained for various stage numbers. Firstly, $R_0$ is set as $5$ while the stage number $N$ is given the values $5, 10, 30$ and $60$, and the corresponding graphs of the solutions are shown in Fig4. In these evaluations, $\epsilon$ and $\gamma_i$ are chosen to be $1/3$ and  $N*0.2$, respectively for  $i=1,...,N$ where $N$ is the infectious stage number.  Fig4 illustrates that just like it is seen in the $SJR$ system, there exists a time shift between the maximum point of $J$ and the inflection point of $R$ in the $SEJR$ model. However, there is a difference between the graphs of the solutions of the classical $SEIR$ model and the model given in (6) since  the $\gamma$ coefficients in the $SEJR$ model change with the stage number $N$. Furthermore, the difference in delay values occurs for relatively small stage numbers.

\begin{figure}[ht]
	\includegraphics[scale=0.57]{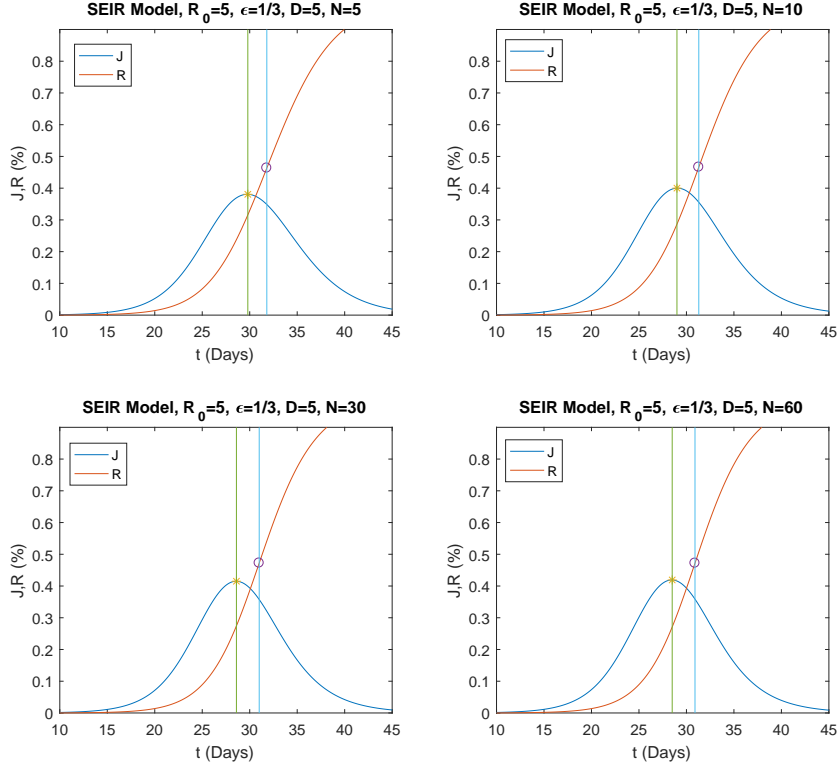}
	\caption{Numerical solutions of multistage SEIR model for different stage numbers $N=5,10, 30$ and $60$. Here $R_0=2.5$, $\epsilon=1/3$ and $\gamma=N$x$0.2$. In the graphs, $*$ represents the location of maximum value of $J$ and $o$ represents the location of the inflection point of $R$. Graphs show that there is a time-shift between these points.  }
\end{figure}

\section{Estimation of the Delay}
\label{Estimation of the delay}

In this section, some theoretical results as well as numerical evaluations on the estimation of the delay between the point where $J$ assumes its maximum value and the inflection point of the removed class $R$ of the model in (5) will be shared. The following proposition shows that  the  value of the difference of the points where each multistage reaches its maximum can be   approximately found.

\textbf{Proposition :} Let $t_i$ be the time where each substage  $I_i$ assumes its maximum and  $1/\gamma_{i}$ be the corresponding infectious period for $i=0,1,...N$. Then quadratic approximation provides that the successive difference  $t_{i}-t_{i-1}$ is   $1/\gamma_{i}$ and hence $t_{i}-t_{j}=\sum_{k=j+1}^{i} \frac {1}{\gamma_{k}}$.
\vskip.2cm
\textbf{Proof :} To determine the distance between the points $t_i$ where each substage  $I_i$  assumes its maximum value we use quadratic approximation of  Taylor series expansion of $I_i$
at the point $t=t_{i-1}$
\begin{eqnarray}
\label{eq:schemeP}
I_i (t)=I_i (t_{i-1})+I_i' (t_{i-1})(t-t_{i-1})+\frac{1}{2} I_i'' (t_{i-1})(t-t_{i-1})^2
\end{eqnarray}
for $i\geq 1$.{ Differentiating (9)
\begin{eqnarray}
\label{eq:schemeP}
I_i' (t)=I_i' (t_{i-1})+I_i'' (t_{i-1})(t-t_{i-1})
\end{eqnarray}
and then substituting $t=t_i$ in (10) and using the fact that  $I_i'(t_i)=0$ since $I_i$ reaches its maximum at $t_i$, one obtains
\begin{eqnarray}
\label{eq:schemeP}
t_i-t_{i-1}=-\frac {I_i' (t_{i-1})}{I_i'' (t_{i-1})}.
\end{eqnarray}
The multistage $SIR$ model defined by the equations in (5) suggests that for $i\geq 1$
\begin{eqnarray}
\label{eq:schemeP}
 I_i'(t)=\gamma_{i-1}I_{i-1}(t)-\gamma_{i}I_{i}(t).
 \end{eqnarray}
Differentiating (12)  yields
\begin{eqnarray}
\label{eq:schemeP}
I_i''(t)=\gamma_{i-1}I_{i-1}'(t)-\gamma_{i}I_{i}'(t).
\end{eqnarray}
The equation (11) together with (13) gives the approximate distance formula as follows
\begin{eqnarray}
\label{eq:schemeP}
t_i-t_{i-1}=\frac {1}{\gamma_i}.
\end{eqnarray}

Proof of the formula for the distance between any $t_i$ is straightforward.

\vskip.3cm
Results obtained by the numerical evaluations   are compatible with the proposition. To observe the distance between the maximum points of the independent  infectious stages, solutions of the multistage $SIR$ model with respect to various infectious periods are chosen. In this respect, the basic reproduction number $R_0$ and  $\gamma_i$ (i.e. duration is 5) are set as $2.5$ and $N*0.2$, respectively and the related graphs of the solutions for various stage numbers  ($N=1,2,3,4$) are given in Fig5. Comparison of graphs in Fig5 reveals that  the  value of the difference of the points where successive stages reach their maximum is approximately $1/\gamma_i$.

\begin{figure}[ht]
	\includegraphics[scale=0.57]{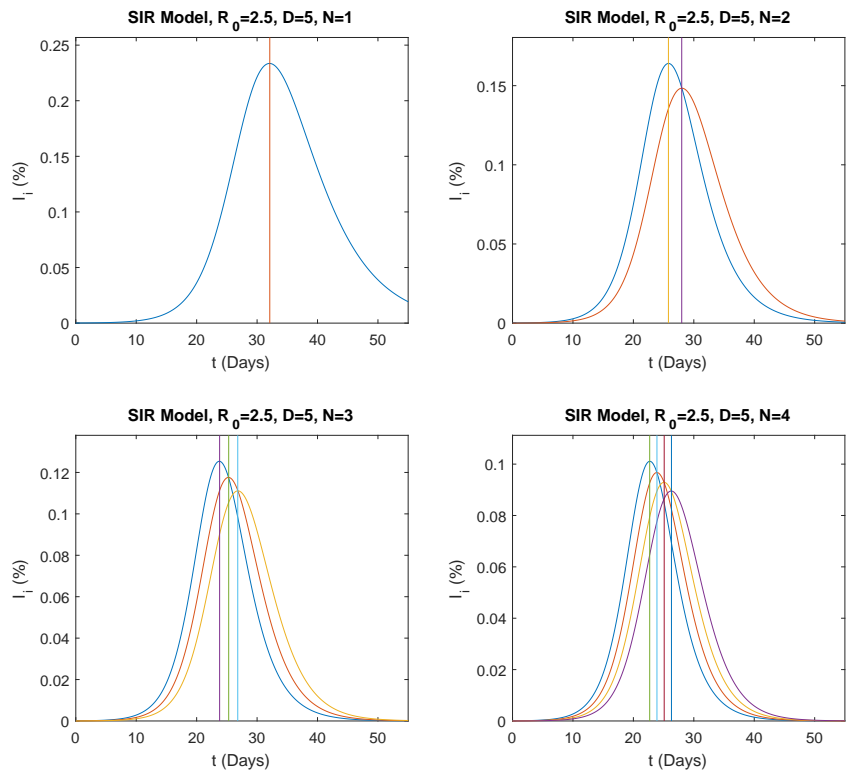}
	\caption{Graphs of the independent infectious stages $I_0,\dots I_N$ from numerical solutions of multistage SIR model (5). Here $R_0=2.5$ and $\gamma=N$x$0.2$. In the graphs, vertical lines indicate the position of maximum points of the independent infectious stages. The distance between these vertical lines is approximately $1/\gamma$.  }
\end{figure}

It should be emphasized that the formula (14) is also valid for the multistage SEIR model. The distance between the maximum points of the independent  infectious stages including the $E$ stage is approximately $1/\gamma_i$. Since the proof is same as in the SIR case, we don't repeat the derivation of formula (14) again to avoid repetition.
However, to observe the distance between the maximum points of the infectious stages numerically,  the basic reproduction number $R_0$, $\epsilon$ and  $\gamma_i$ (i.e. duration is 5) are set as $5$, $1/3$ and $N$x$0.2$. Then, the $SEIR$ model is solved for $N=1,2,3$ and $4$, and the related graphs are given in Fig6. As in the case of the $SIR$ model, it is observed that the distance between the maxima is found approximately  $1/\gamma_i$, too.}
\begin{figure}[ht]
	\includegraphics[scale=0.57]{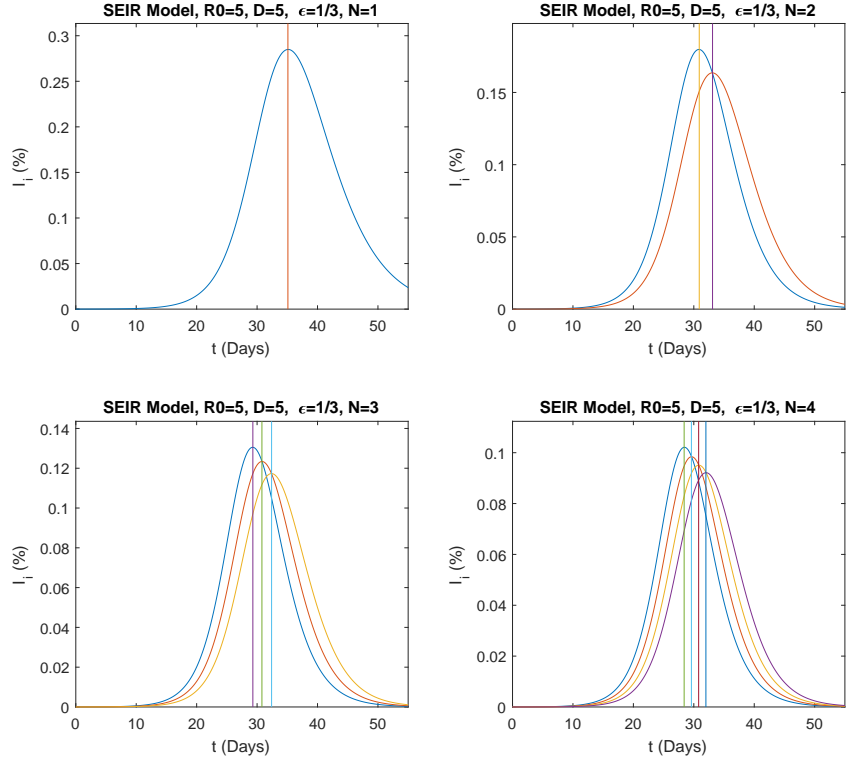}
	\caption{Graphs of the independent infectious stages $I_1,\dots I_N$ from numerical solutions of multistage SEIR model (6). Here $R_0=2.5$, $\epsilon=1/3$ and $\gamma=N$x$0.2$. In the graphs, vertical lines indicate the position of maximum points of the independent infectious stages. The distance between these vertical lines is approximately $1/\gamma$.}
\end{figure}

\section{Numerical Results for Models with Multiple Infectious Stages}
\label{Numerical Results for Models with Multiple infectious Stages}

In the previous section, it is shown theoretically that the multistage infectious $SIR$ model defined  by the equations in (5) explains the epidemiological delay dynamic between the infectious class $I$ and the removed class $R$. In this section, the numerical evaluations of the system (5) for some infectious models are obtained and these observations are  used to provide some compatible results of the theoretical findings.

\begin{figure}[ht]
\includegraphics[scale=0.57]{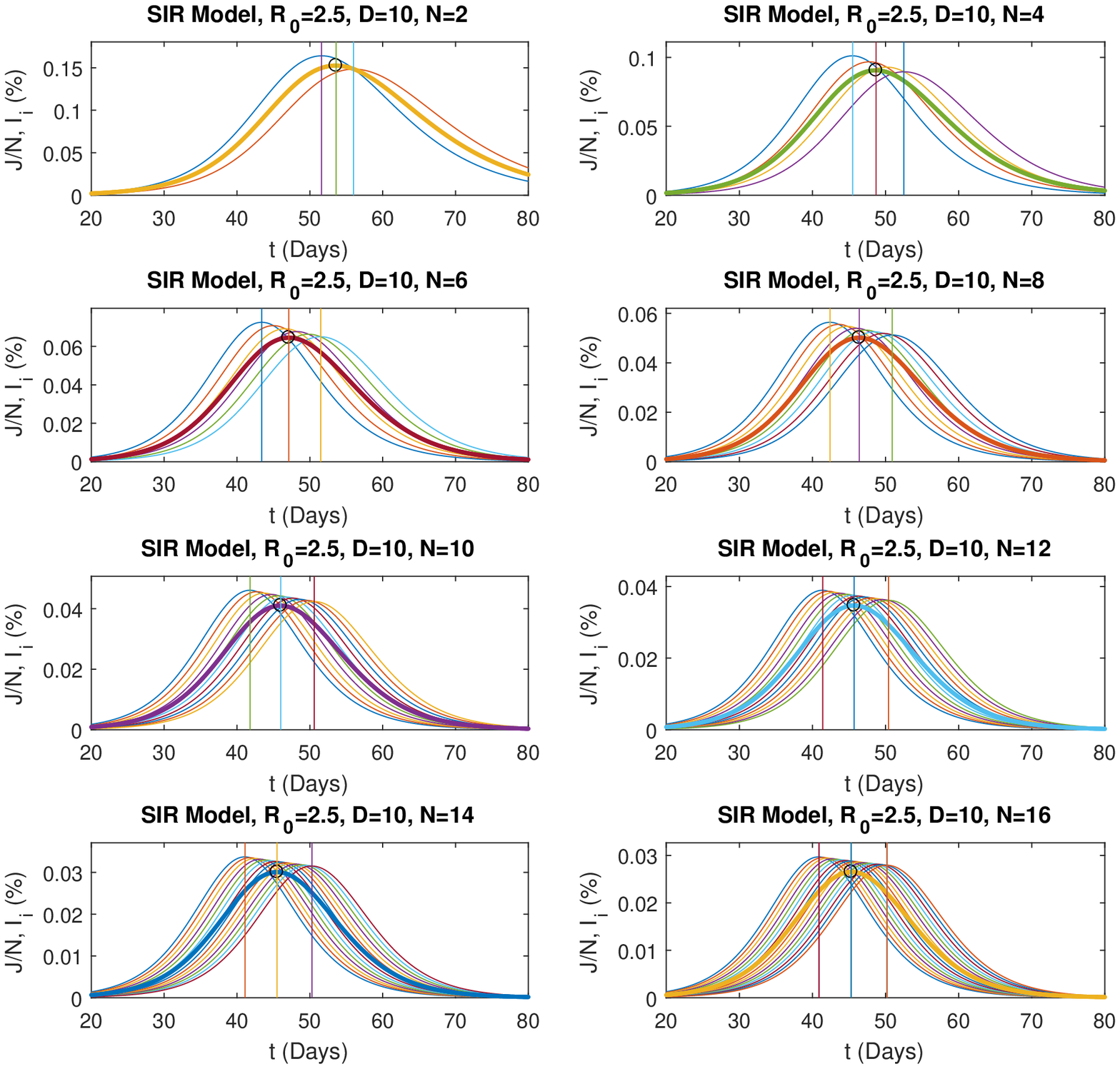}
\caption{Change of the position of maxiumum of $J/N$ (normalized $J$) in time with respect to number of stages $N$ for multistage SIR model."o"  indicates the position of the  maxima of $J/N$ (thick curves). Vertical lines represent the position of maximum points of the first stage,  $J/N$ and the last stage, respectively. }
\end{figure}

To analyze the behaviour of the time shift for relatively large values of stage number, the graphs of the solutions for the same epidemic parameters are obtained for different stage numbers $N$, and the corresponding graphs are given in Fig7. As it is clear in Fig7, the solution curves remain unchanged as the stage number $N$ increases.

To investigate the effect of the basic reproduction number $R_0$ and the infectious period $1/\eta$ on the infectious dynamics and the resulting delay, the system (5) is solved with the initial conditions given by  (7) for some parameter values. To this end, the pair  $(R_0, 1/\eta)$ is chosen $(2.5,5)$, $(5,5)$, $(10,5)$, $(2.5,10)$, $(5,10)$, $(10,10)$, $(2.5,20)$, $(5,20)$ and $(10,20)$, respectively and the numerical evaluations for various  infectious stages  $N$ are shown in Fig 8.  It can be observed from this figure that the delay is almost half of the infectious period.  The compatibility can also be seen in Fig. 7 where the the time value of the maximum of $J/N$  (normalized $J$) in time is located at the middle of time values of the maxiumum of first infectious stage and the maximum of last infectious stage.

\begin{figure}[ht]
	{
		$\hspace*{-1cm}\begin{array}{ccc}
		\includegraphics[scale=0.64]{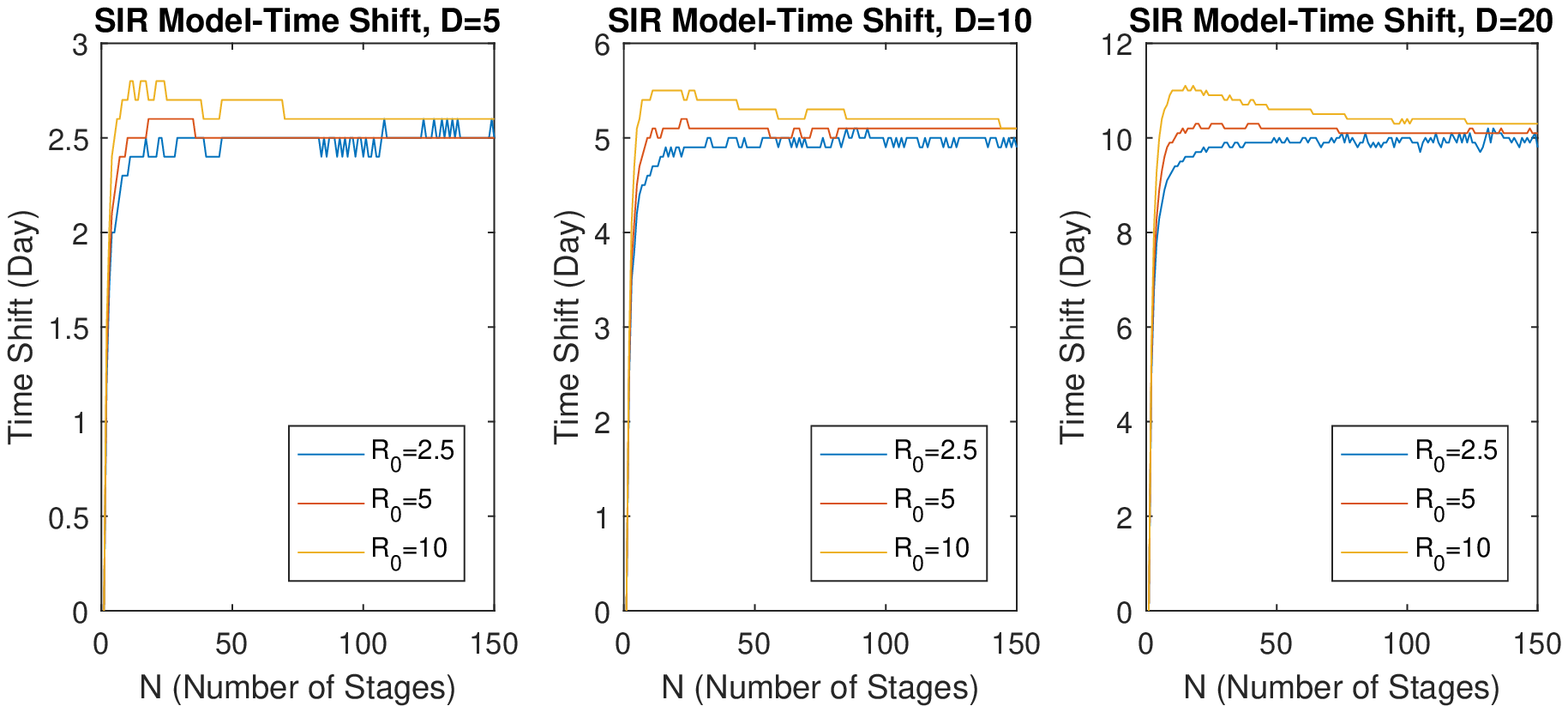}
		\\ \rm (a)\hspace{3cm} \rm (b)\hspace{3cm}   \rm(c) \\
		\end{array}$}
	{\small
		\caption{Variations of time shift (delay) with resprect to number of stages $N$ for $R_0=2.5,\,5$ and $10$; $D=5,\,10$ and $20$ in the multistage SIR model. For all cases the time shift become stable at a constant value after a critical stage number $N$.}
		\label{figure-30derivatives}}
\end{figure}

Comparison of panels of Fig. 8 shows that the change in the reproduction number  $R_0$ for a fixed infection period has no effect on the delay time  $d_t$. Moreover, the infection period affects the change in  $d_t$ value which is nearly half of the infection period. One of the most important characteristics of the Fig. 7 and Fig.8  is that the multistage SIR model (5) offers a more realistic explanation for the delay in the epidemiological dynamics.

To analyze the behaviour of the multistage $SEIR$ model for relatively large values of $N$,  the system defined by (6) is solved for different stage numbers $N$, and for the same epidemic parameters chosen above, and the related curves are given in  Fig 9 which illustrates that the solution curves of the multistage $SEIR$ model remain unchanged as $N$ increases.

\begin{figure}[ht]
\includegraphics[scale=0.57]{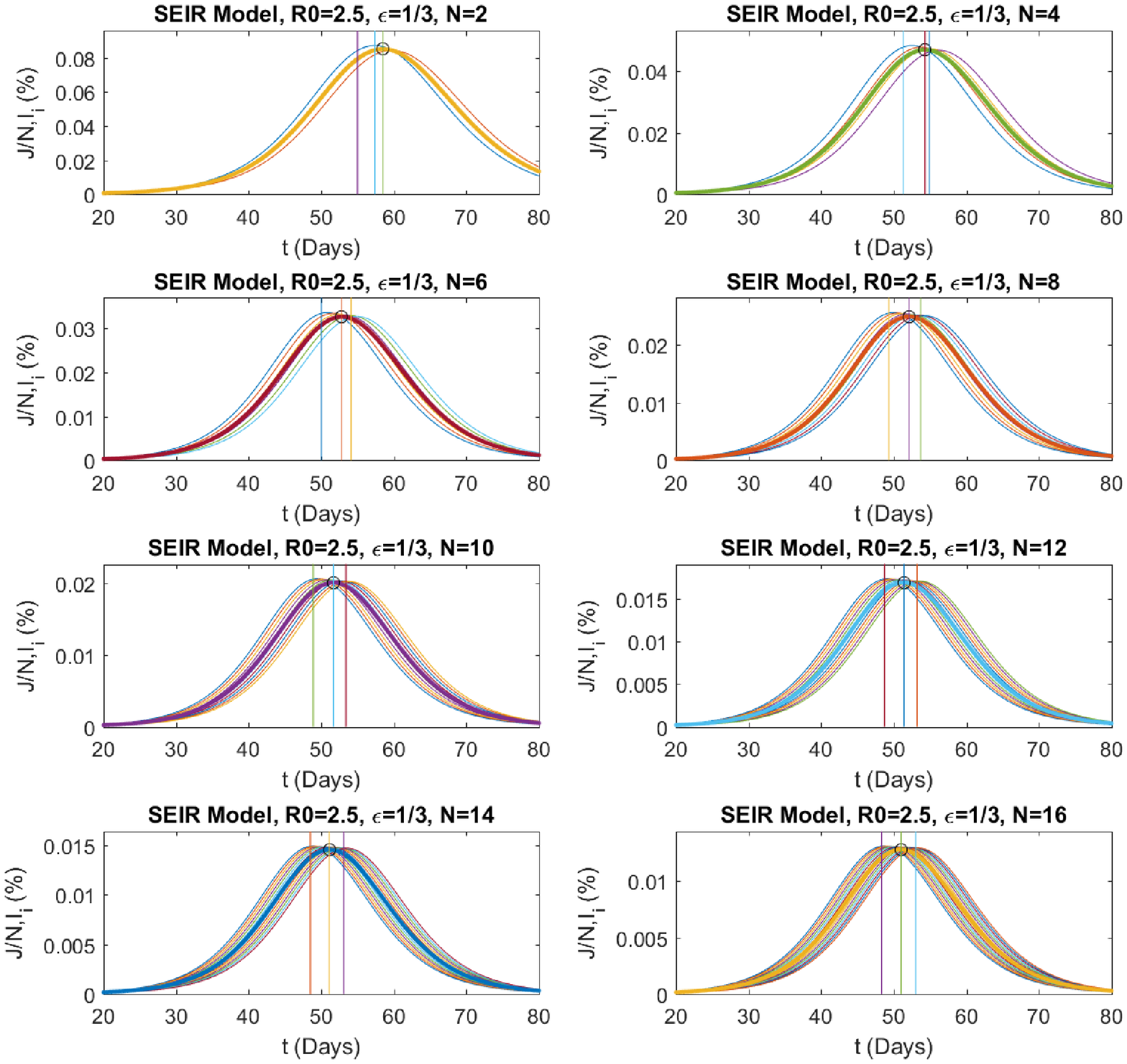}
\caption{Change of the position of maxiumum of $J/N$ (normalized $J$) in time with respect to number of stages $N$ for multistage SEIR model. "o"  indicates the position of the  maxima of $J/N$ (thick curves). Vertical lines represent the position of maximum points of the first stage,  $J/N$ and the last stage, respectively.  }
\end{figure}

\begin{figure}[ht]
\includegraphics[scale=0.57]{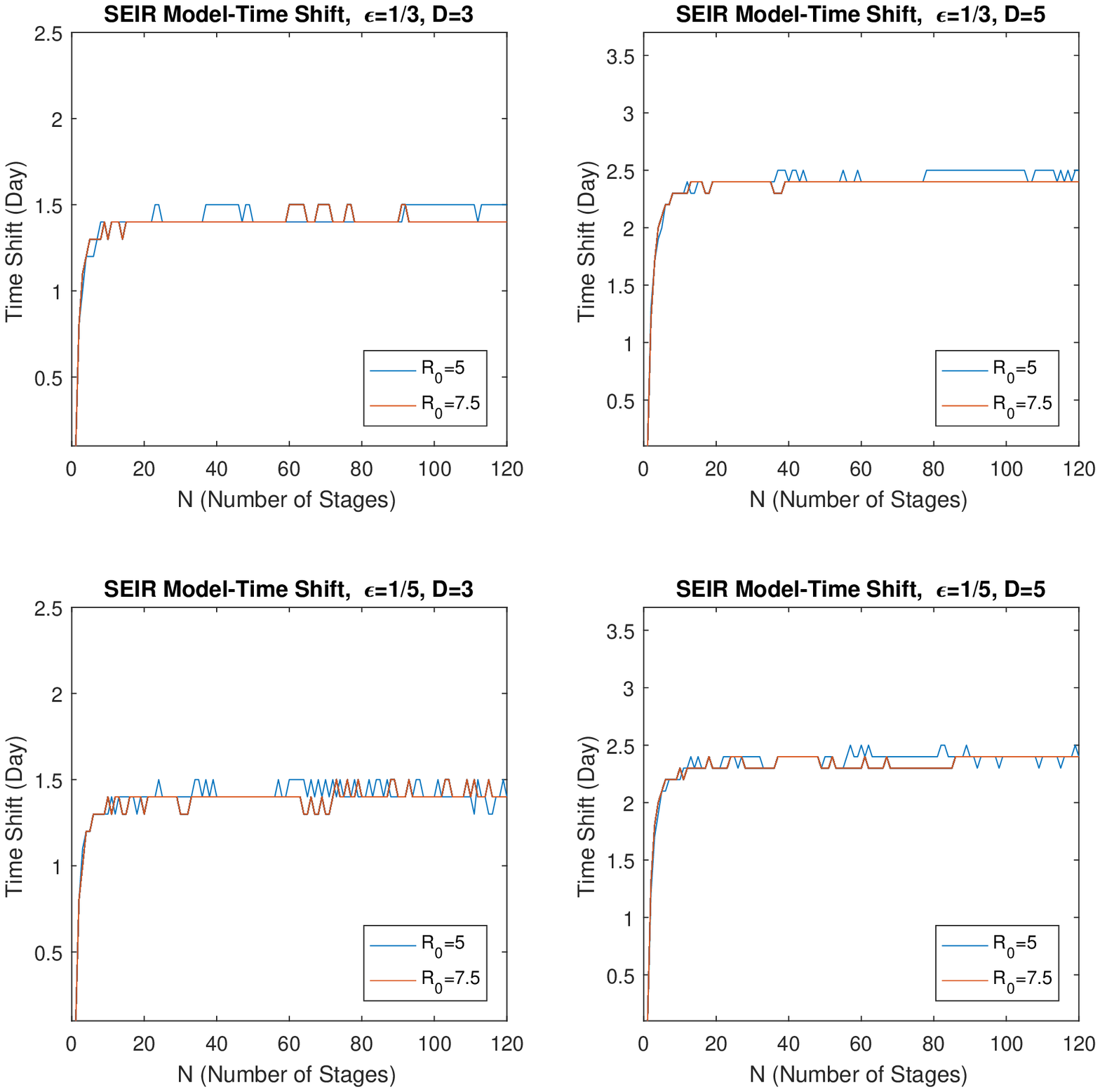}
\caption{Variation of time shift for $R_0=5$ and $7.5$ in the multistage SEIR model. Here $\epsilon=1/3,\,1/5$ and $D=3,\,5$ are taken. For each case, delay becomes stable at a constant value after a critical stage number $N$.}
\end{figure}

Similar analysis can be performed to the  multistage $SEIR$ model. For the observation of the change in the delay value due to the stage number for the multistage $SEIR$ model,  $\epsilon$ is chosen to be $(1/3)$ and  $(1/5)$ whereas  $\gamma_i$ is set as $N/3$ and $N/5$. Then, the graphs that show the change in the delay due to the value of $N$ for  $R_0=5$ and $R_0=7.5$ are given in  Fig 10. An analysis of these graphs yields that as $N$  increases, the delay converges to a value. Moreover, it could easily be observed that the delay is independent of  $\epsilon$ and $R_0$ but yet it is influenced by the values of  $\gamma_i$ (i.e. infectious period). Even though the delay is smaller that the delay observed in the  multistage $SIR$ model, it is still approximately half of the infectious period. The compatibility can also be seen in Fig9 where the the time value of the maximum of $J/N$ is located almost at the middle of time values of the maximum of first infectious stage and the maximum of last infectious stage.

\section{Conclusion}
\label{Conclusion}

Epidemic models based on the real data signify that there exists a time shift between the time value for which the infectious stage reaches its maximum  and the inflection point of  the recovered curve. However, the epidemic models with a single infectious stage ($SIR$ and  $SEIR$) can not project this delay.  In this article, we propose the multistage versions of these models to represent this time shift.

In accordance with this purpose, the differential equation systems of these multistage models are evaluated for specifically chosen epidemic parameters.  When the solutions of these systems are analyzed, it is seen that the  multistage models successfully reveal the time shift. While the delay varies  for relatively small stage numbers, it is observed that the delay becomes nearly stable as $N$  increases.  Furthermore, it is also observed that the basic reproduction number does not affect the delay value whereas the delay changes due to the  infectious period.

Additionally,  the distance between the points where each infectious stage reaches its maximum is  found approximately  $1/\gamma_i$ both graphically and qualitatively. Numerical solutions of  these multistage models of epidemic diseases show strong evidences that the delay is approximately half of infection period of the disease.

As a result, it is shown that the delay phenomenon observed in the infectious diseases defined by the epidemic models  $SIR$ and  $SEIR$ can be successfully explained by the multistage forms of these models.

\clearpage
\newpage
\section*{References}
\bibliographystyle{elsarticle-num}
\biboptions{compress}

\end{document}